# Electron Spin Dynamics of the Intersystem Crossing in Aminoanthraquinone Derivatives: The Spectral Telltale of Short Triplet Excited States

*Ruilei Wang,[a]† Andrey A. Sukhanov,[b]†Yue He,[c]† Aidar Mambetov,[b] Jianzhang Zhao,[a]\* Daniel Escudero,[c]\* Violeta K. Voronkova [b,]\* and Mariangela Di Donato[d,e]\**

[a]State Key Laboratory of Fine Chemicals, Frontiers Science Center for Smart Materials, School of Chemical Engineering, Dalian University of Technology, 2 Ling Gong Road, Dalian 116024, P. R. China. *Email: zhaojzh@dlut.edu.cn (J.Z.)

[b]Zavoisky Physical-Technical Institute, FRC Kazan Scientific Center of Russian Academy of Sciences, Kazan 420029, Russia. *E-mail: vor18@yandex.ru (V. K V.)

[c] Department of Chemistry, KU Leuven, Celestijnenlaan 200F, 3001 Leuven, Belgium
E-mail: Daniel.Escudero@kuleuven.be (D.E.)

[d]LENS (European Laboratory for Non-Linear Spectroscopy) via N. Carrara 1, 50019 Sesto Fiorentino (FI), Firenze, Italy. *E-mail: didonato@lens.unifi.it (M. D.D.)

[e]ICCOM-CNR, via Madonna del Piano 10, 50019 Sesto Fiorentino (FI), Italy

† These authors contributed equally to this work.

**ABSTRACT:** Herein we studied the excited state dynamics of two bis-amino substituted anthraquinone (AQ) derivatives. Femtosecond transient absorption spectra show that intersystem crossing (ISC) takes place in 190 – 320 ps, and nanosecond transient absorption spectra demonstrated unusually short triplet state lifetime (2.06–5.43 µs) for the two AQ derivatives at room temperature. Pulsed laser excited time–resolved electron paramagnetic resonance (TREPR) spectra shows an inversion of the electron spin polarization (ESP) phase pattern of the triplet state at longer delay–time. Spectral simulations show that the faster decay of the $T_y$ sublevel ($\tau_x$ = 15.0 µs, $\tau_y$ = 1.50 µs, $\tau_z$ = 15.0 µs) rationalizes the short $T_1$ state lifetime and the ESP inversion. Computations taking into account the electron-vibrational coupling, i.e. the Herzberg–Teller effect, successfully rationalize the TREPR experimental observations.

**TOC GRAPHICS**

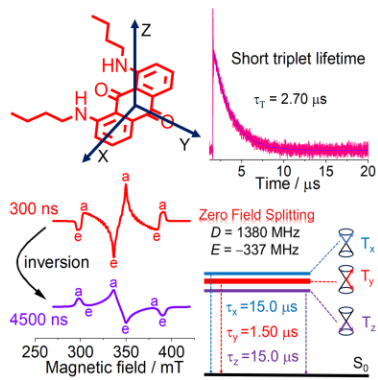

Triplet photosensitizers (PSs) have attracted much attention during recent years, both in fundamental photochemistry studies,[1,2] as well as for applications in photocatalytic organic synthesis,[3–5] photocatalytic $H_2$ production or $CO_2$ reduction,[6,7] photodynamic therapy (PDT),[2,8–10] or triplet–triplet annihilation (TTA) photon upconversion.[11–15] Triplet PSs show strong absorption of light, efficient intersystem crossing (ISC) upon photoexcitation, appropriate excited state redox potentials, and good thermal- and photo-stability.[16] Among these properties, the triplet excited state lifetime is a critical parameter for their applications, which often require intermolecular energy or electron transfer from the triplet excited state of the photosensitizer to the substrates.[17] Longer triplet state lifetime lead to enhanced yields of triplet energy or electron transfer,[18–23] being the latter processes usually diffusion–controlled. However, controlling the triplet excited state lifetimes through structural design is still a major challenge in photochemistry. Since closely related chromophores may display drastically different triplet state lifetimes, in-depth investigations are needed to disclose the relationship between the molecular structures and the triplet state lifetimes.[24,25]

The typical optical spectroscopic methods to study the triplet excited states of organic chromophores include steady state and time–resolved transient absorption and luminescence spectroscopy.[26–28] The phosphorescence of the emissive $T_1$ state can be observed (either at room temperature in fluid solution or in frozen solution at low temperature), and the phosphorescence lifetimes can be measured. Based on these data along with the phosphorescence quantum yield, the radiative and non–radiative decay rate constants from the $T_1$ state can be determined. On the other hand, the femtosecond/picosecond transient absorption spectra can supply information about the ISC kinetics, and in some cases, the ISC mechanism, revealing for instance, charge separation (CS) and recombination (CR) induced ISC, or the spin–orbit coupling (SOC) induced ISC in heavy atom–containing molecules.[24,29–31] Finally, nanosecond transient absorption spectroscopy can give information about the triplet state lifetimes (usually in the range of microseconds, beyond the detection window of the femtosecond transient absorption spectroscopy).[32–35]

In all these ordinary optical spectroscopic methods, the $T_1$ state is treated as one single electronic state. However, it is known that in molecules with low symmetry, because of the effect of the zero field splitting (ZFS), which originates from the the loss of the degeneracy of the spin components of a spin-orbit free triplet state, three sublevels exist for the $T_1$ state, i.e. $T_x$, $T_y$ and $T_z$. These sublevels are non-degenerate, although the energy difference between them can be as low as ~ 0.01 cm$^{-1}$ or even smaller.[36] Nevertheless, the three spin sublevels may show different properties, as for instance, different lifetimes, which can affect the overall triplet lifetime given the transition between the triplet sublevels is fast.[37–39] Unfortunately, it is challenging to characterize the three sublevels with the above mentioned optical spectroscopic methods. High resolution phosphorescence spectroscopy based on the Shpol'skii matrix technique has been used to obtain the energy gap between the sublevels and their respective lifetimes,[37–40] however, this method is technically challenging, and requires the $T_1$ state to be phosphorescent. Moreover, this method does not give information about the electron spin polarization (ESP) of the triplet state, i.e. the population ratios of the $T_x$, $T_y$ and $T_z$ sublevels of the $T_1$ state. SOC is normally anisotropic thus resulting in different ISC rate constants for the $S_1 \rightarrow T_x$, $S_1 \rightarrow T_y$ and $S_1 \rightarrow T_z$ channels. As a result, the population ratios of the three sublevels of the $T_m$ state are severely deviated from the Boltzmann distribution. Optically detection magnetic resonance (ODMR) spectra can characterize the energy difference of the three sublevels of the $T_1$ state, but they can hardly directly supply any information about their lifetimes and their population ratios. Concerning the above challenges, pulsed laser excited continuous wave (CW) time–resolved electron paramagnetic resonance (TREPR) spectroscopy is of particular interest.[36,41–43] The transient paramagnetic species, such as triplet states, can be recorded with TREPR spectroscopy and the electron spin polarization can be also characterized. Furthermore, the ZFS *D* and *E* parameters are feasibly derived from the TREPR spectral data, whose simulation gives information about the population ratios of the three spin sublevels.[41]

We have prepared the two anthraquinone derivatives (Scheme 1) which absorbe in the visible region of the UV-vis spectrum. These derivatives have been used as photoinitiators for photopolymerization.[44] Their triplet excited state lifetimes are unusually short (2.06–5.43 μs. Table 1), and we hypothesized that one sublevel of the $T_1$ state may have a very short lifetime.[25,45–48] The optical characterization of the compounds did not provide any further insight about these aspects. Therefore, we use TREPR spectroscopy to characterize the $T_1$ excited state of the compounds. Indeed, we found that one sublevel of the $T_1$ state decays 10–times faster than the other two sublevels. We propose this is probably a general property for triplet PSs showing short triplet excited state lifetime.[25,49,50]

**Scheme 1. Molecular Structures of the Compounds Used in Study**[a]

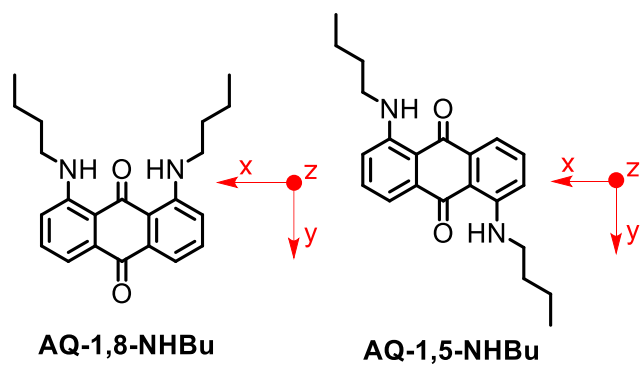

[a] Molecular structures of the AQ derivatives used in the studies. The orientation of the ZFS principal directions (*X*, *Y* and *Z*) of $T_1$ state are shown for compounds of **AQ-1,8-NHBu** and **AQ-1,5-NHBu**.

Anthraquinone (AQ) is known for its fast and efficient ISC.[51] However, the pristine AQ is not an ideal triplet PS to be used with visible light excitation, because it only absorbs in the UV/blue range. To shift its absorption, we attached an electron donating amino moiety to the AQ chromophore. The absorption of the resulted derivatives is red−shifted towards the green range (*ca.* 524 nm) (Figure 1 and Table 1) compared to bare AQ chromophore (*ca.* 325 nm).[33] We prepared two analogues, with the *n*-butylamino substituents attached at 1,8- and 1,5-positions, respectively (**AQ-1,8-NHBu** and **AQ-1,5-NHBu**, Scheme 1).

The two compounds show similar absorption and fluorescence bands, with the emission maximum peaking at *ca.* 580 nm (Figure 1). The maximum fluorescence emission wavelength and their intensities are dependent on the solvent polarity, leading to weaker and red-shifted emission maxima with increasing solvent polarity.[44] Interestingly, we observed significant ISC for these heavy atom−free compounds. The singlet oxygen quantum yield ($\Phi_\Delta$) is up to 33.3% (Table 1). We found fast ISC for the compounds (Figure 2). ISC rate constant calculations were also performed (Table 3 and Table S2).

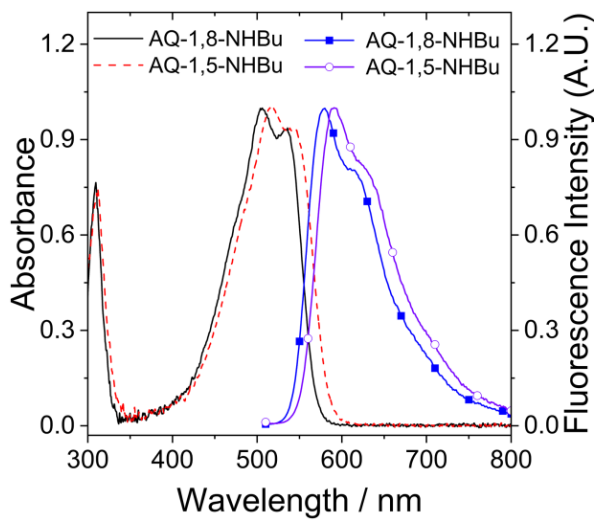

**Figure 1.** The normalized UV−vis absorption (black and red line) ($c = 1.0 \times 10^{-5}$ M) and fluorescence emission spectra (blue and purple line) ($A = 0.1$ at $\lambda_{ex} = 500$ nm) of the compounds in toluene (TOL). 20 °C.

The triplet states of the compounds have been studied with nanosecond transient absorption spectroscopy,[44] and the lifetimes are collected in Table 1. The lifetimes of the triplet states were determined to be 2.06−5.43 μs (Table 1), respectively for **AQ-1,8-NHBu** and **AQ-1,5-NHBu**; which are much shorter than that expected for a heavy atom−free organic chromophore.[16,17] However, these triplet state lifetimes are close to the AQ derivatives substituted with chloride atoms.[52] The lifetimes are much shorter than the AQ derivatives with carbonyl groups attached to the chromophore (the triplet state lifetime is *ca.* 30 μs).[51]

**Table 1. Photophysical Parameters of the Compounds.**

|  | Solvents | $\lambda_{abs}^a$ / nm | $\varepsilon^b$ | $\lambda_{em}^c$ / nm | $\Phi_\Delta^d$ / % | $\tau_T^e$ / μs |
|---|---|---|---|---|---|---|
| **AQ-1,8-NHBu** | CHX | 506 | 1.5 | 571 | 24.5 | 2.06 |
|  | TOL | 506 | 1.4 | 580 | 32.8 | 2.70 |
|  | DCM | 522 | 1.5 | 605 | 30.9 | 2.81 |
|  | ACN | 520 | 1.4 | 617 | 17.8 | 3.74 |
|  | MeOH | 524 | 1.4 | 621 | 6.7 | 5.43 |
| **AQ-1,5-NHBu** | CHX | 506 | 1.7 | 574 | 20.9 | 2.31 |
|  | TOL | 517 | 1.5 | 591 | 26.7 | 2.40 |
|  | DCM | 525 | 1.6 | 604 | 33.3 | 2.96 |
|  | ACN | 519 | 1.6 | 617 | 19.6 | 3.72 |
|  | MeOH | 524 | 1.6 | 621 | 6.4 | 5.37 |

$^a$ Maximal UV–vis absorption wavelength, $c$ =1.0 × 10$^{-5}$ M, 20 °C. $^b$ Molar absorption coefficient at absorption wavelength, $\varepsilon$: 10$^4$ M$^{-1}$ cm$^{-1}$. $^c$ Fluorescence emission wavelength, $\lambda_{ex}$ = 500 nm. $^d$ Singlet oxygen ($^1O_2$) quantum yield with 2,6-diiodo-Bodipy as standard ($\Phi_\Delta$ = 87% in DCM), $\lambda_{ex}$ = 500 nm. $^e$ Triplet lifetimes with different polarity solvents.

To elucidate the dynamics of the intersystem crossing of the compounds, we measured the femtosecond transient absorption spectra of **AQ-1,8-NHBu** and **AQ-1,5-NHBu** in solvents with different polarity (Figure 2). All the spectra have been measured by exciting the samples at 490 nm and the data were analyzed with singular value decomposition (SVD) and global fitting,[28,53] using a sequential decay scheme. Figure 2a and 2b show the Evolution Associated Difference Spectra (EADS) obtained from the analysis of the transient data measured for **AQ-1,8-NHBu** and **AQ-1,5-NHBu** in cyclohexane (CHX).

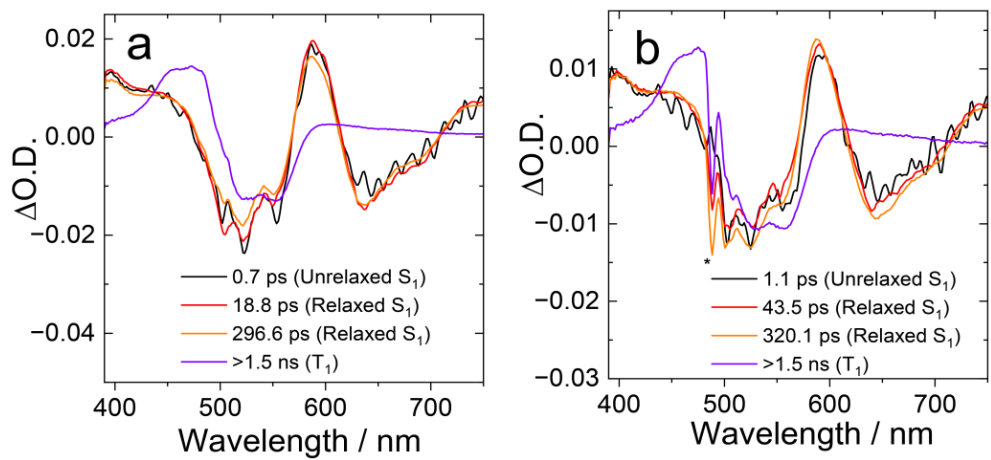

**Figure 2.** Femtosecond transient absorption spectra of the compounds, the Evolution Associated Difference Spectra (EADS) obtained from global fitting analysis for (a) **AQ-1,8-NHBu** and (b) **AQ-1,5-NHBu**. * represents the scattering from the pump beam. $\lambda_{ex}$ = 490 nm, in cyclohexane (CHX), 20°C.

At short time scale, the transient spectra of **AQ-1,8-NHBu** in CHX (Figure 2a) present two negative signals, respectively peaked at about 525 and 635 nm, assigned to ground state bleaching (GSB, 525 nm) and stimulated emission (SE, 635 nm) and two excited state absorption bands, a broad one extending at wavelengths < 470 nm, and a sharper one peaked at about 590 nm, which are assigned to the $S_1 \rightarrow S_n$ transition based on previous studies.[54] Up to about 18.8 ps, the dynamic evolution observed for **AQ-1,8-NHBu** can be attributed to vibrational and solvent induced relaxation of the $S_1$ state. A comparison with previous results reported in the literature,[44] reveals that the long living spectral component (purple line) likely corresponds to the triplet state. The ISC rate for the **AQ-1,8-NHBu** is thus determined to be 296.6 ps (Figure 2a). Transient spectra recorded in different solvents are similar to those obtained in CHX, and the ISC rate of **AQ-1,8-NHBu** shows a limited dependence on the solvent polarity (Table S1). The transient absorption features of **AQ-1,5-NHBu** are very similar to those of **AQ-1,8-NHBu** (Figure 2b and Table S1). We note that in comparison to AQ derivatives containing no amino substituents, the ISC kinetics of **AQ-1,5-NHBu** and **AQ-1,8-NHBu** are much slower (*ca.* 0.4 ps for the later compounds).[51]

Besides the dynamics of the electronic states, the electron spin polarization of the triplet states offer rich and unique information about the electron spin dynamics of the compounds upon photoexcitation, and the zero field splitting (ZFS) of the triplet state provides information about the triplet state wave function, for instance about its electronic configuration (whether a $^3$n-π* state or a $^3$π-π* state), or about the spatial confinement of the triplet state wave function.[25,55–59] Time-resolved electron paramagnetic resonance (TRERP) spectroscopy is a dedicated method for these measurements.[41,42,60] The TREPR spectra of the compounds in frozen solution at low temperature were recorded (Figure S3 and Figure S4). For **AQ-1,8-NHBu**, the triplet state TREPR spectra shows ESP at the six canonical orientation as (e,a,e,a,e,a) (e: emissive; a: enhanced absorptive) (Figure 3a). Note that this is not the typical ESP pattern for a triplet state formed via the spin orbital coupling (SOC) effect, which usually shows an ESP phase pattern as (e,e,e,a,a,a) or (a,a,a,e,e,e) (in plane population).[36,41] The experimental TREPR spectra and their time–evolution were simulated by using a program simulating EPR line shape of triplets with random distribution and non-Boltzmann population distribution of the spin levels, taking into account zero field splitting (ZFS), the anisotropy of the decay rates of sublevels of the triplet state, spin–lattice and spin–spin relaxations.[61] The ZFS $D$ and $E$ paremetersare determined 1380 MHz and −337 MHz, respectively (Table 2) for **AQ-1,8-NHBu**. The population ratios of the three spin sublevels of the $T_1$ state are determined as $P_x : P_y : P_z$ = 0 : 0.9 : 0.1 (Table 2). This result indicates that the $T_y$ sublevel of the $T_1$ state is over populated as compared to the other two spin sublevels. Similar results were observed for **AQ-1,5-NHBu** (Table 2 and Figure 3b). In addition, the theoretically calculated ZFS, $D$ and $E$ parameters of the two compounds are also in good agreement with the experimental results (Table S5).

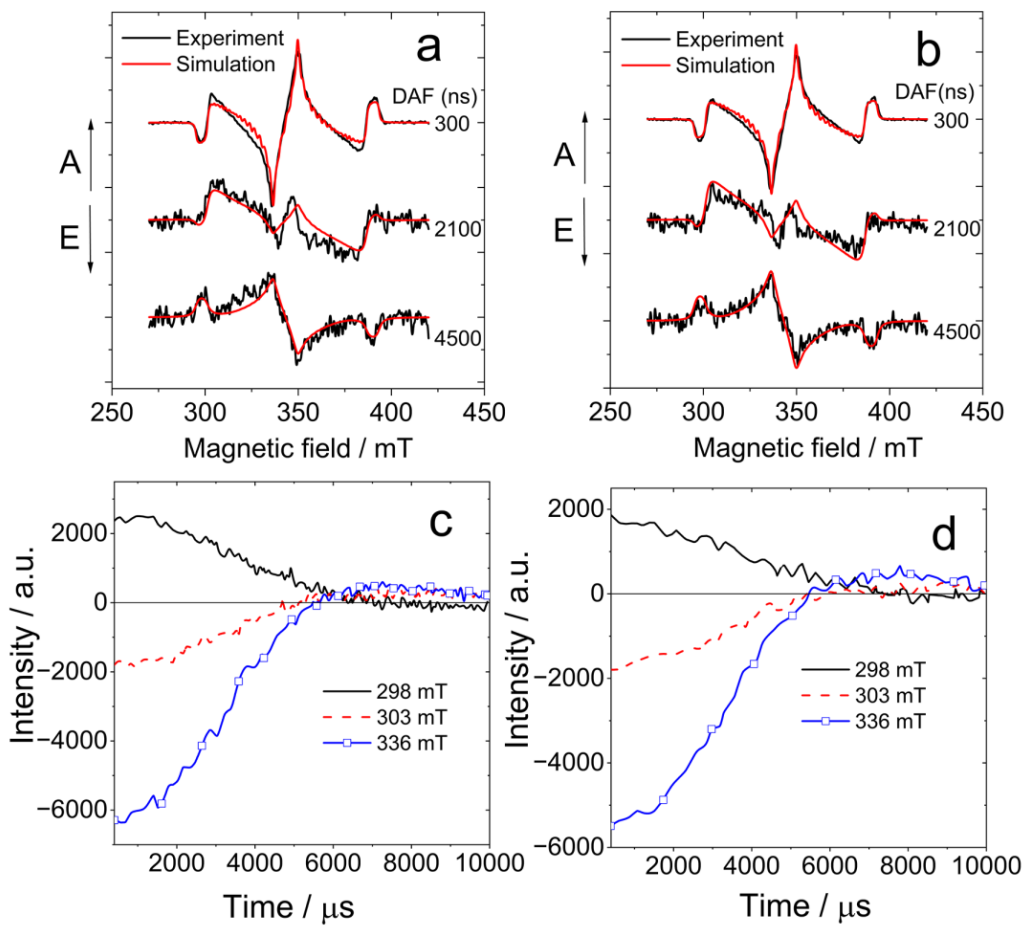

**Figure 3.** TREPR spectra of (a) **AQ-1,8-NHBu** and (b) **AQ-1,5-NHBu** at X-band, in frozen solution at 80 K. Determined with X-band EPR spectromete. The red lines are computer simulations (with EasySpin Programe) of the triplet-state spectra with parameters supplied in Table 2. The corresponding evolution of the signal at specific magnetic field decay traces are presented in (c) for **AQ-1,8-NHBu** and in (d) for **AQ-1,5-NHBu**. The spectra were recorded after 300 ns following a 506 nm laser pulse, $c = 1.0 \times 10^{-4}$ M in mixed solvent of TOL/2-MeTHF (1/3, v/v). A positive sign corresponds to the enhanced absorption signal (A); a negative sign corresponds to the emission signal (E).

Interestingly, at longer delay times after the laser flash, we found that the ESP phase pattern inverts for **AQ-1,8-NHBu** (Figure 3a and Figure 3c), becoming (a,e,a,e,a,e), as shown in Scheme 2. Since the ZFS *D* and *E* did not change during the evolution of the TREPR spectra, the inversion of the ESP phase pattern is unlikely caused by photo decomposition. Similar results were observed for **AQ-1,5-NHBu** (Figure 3b and Figure 3d). This result indicates that one sublevel of the $T_1$ state decays much faster than the other two (Table 2).

Previously, ESP inversion was observed for the TREPR spectra of the triplet states of pyridazine,[46] $C_{60}$ and $C_{70}$ at 3 K.[62] In these molecular system, the ESP inversion was attributed to the fact that the population and the selective decay rate constants of the three sublevels of $T_1$ state are anisotropic. In addition, ESP inversion of tropone at 4.2 K and 2-cyclopentenone at 77 K were also observed.[48,63] In these molecular systems, the ESP inversion was also attributed to the different decay rates of the triplet sublevels of the $T_1$ state. Recently, our research group found ESP inversion in some chromophore–radical dyads at room temperature, for example, in case of the bay-substituted perylenebisimide–TEMPO (PBI-TEMPO) dyad. In this case, ESP inversion was attributed to the initial completion of the quartet state population and the subsequent depopulation process of doublet state induced by the ZFS.[64]

ESP inversion in the TREPR spectra of the triplet states of 2,6-diiododistyryl BODIPY was also observed, which indicated a 37–fold different decay rates for the $T_x$, $T_y$ and $T_z$ sublevels of the $T_1$ state. Simulation of the time–dependent TREPR spectra with our program,[61] show that one spin sublevel of their $T_1$ states decays much faster than the other two sublevels for $D > 0$ and the decay lifetimes of the three sublevels are $\tau_x = 15.0$ μs, $\tau_y = 1.50$ μs and $\tau_z = 15.0$ μs, respectively (Table 2). Note the decay of the $T_y$ sublevel is 10 times faster than the other two sublevels (Table 2). These results indicate that a specific sublevel may impose a substantial effect on the lifetime of the $T_1$ state.[38,39] However, some prerequisites should be satisfied for observation of such phenomenon, i.e. if the triplet lifetime is short, then even if the sublevels have different lifetimes, the inversion may not be observed, since the EPR spectrum will decay before the population inversion appears. It all depends on the relationship between the spin-lattice relaxation time, the triplet lifetime and the difference in lifetime for different sublevels. These results offer in-depth understanding of the triplet excited state lifetime, as well as for improvement their application in OLED (for instance, to shorten the phosphorescence, thus to avoid the efficiency roll-off effect),[38,39] and PDT,[25,58] photocatalysis,[65–67] etc. In the later two cases, longer triplet state lifetime is desired to ensure efficient intermolecular electron transfer or enerny transfer.

**Table 2. Zero Field Splitting Parameters ($D$ and $E$), Relative Population Rates $P_x$, $P_y$, $P_z$ and Lifetimes $\tau_x$, $\tau_y$ $\tau_z$ of the Triplet Spin Sublevels Obtained from Simulations of the TREPR Spectra of the Compounds** [a]

|  | $D^b$ | $E^b$ | $P_x$ | $P_y$ | $P_z$ | $\tau_x^c$ | $\tau_y^c$ | $\tau_z^c$ |
|---|---|---|---|---|---|---|---|---|
| **AQ-1,8-NHBu** | 1380 | −337 | 0 | 0.9 | 0.1 | 15.0 | 1.50 | 15.0 |
| **AQ-1,5-NHBu** | 1380 | −337 | 0 | 0.9 | 0.1 | 15.0 | 1.50 | 15.0 |

[a] Obtained from simulation of the TREPR spectra of the compounds in the mixed solvent of TOL/2-MeTHF (1/3, v/v), T = 80 K. [b] In MHZ. [c] In μs.

**Scheme 2. Simplified diagram explaining the inversion of electron spin polarization of AQ-1,8-NHBu and AQ-1,5-NHBu** [a]

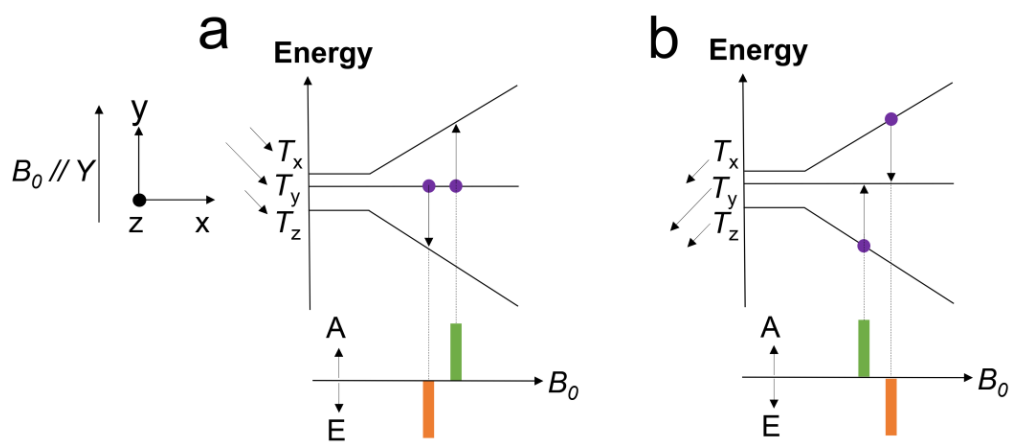

[a] (a) Just after the laser excitation, $T_y$ sublevel of the $T_1$ state is over populated and mainly produces an **E/A** pattern. (b) Sufficiently later (longer delay time), the rapid decay rate of the $T_y$ sublevel of the $T_1$ state makes the population of the $T_y$ sublevel deplete selectively and the observed polarity becomes an **A/E** pattern.

The electronic configuration of the $T_1$ state was determined from the ZFS parameters obtained from the analysis of the TREPR spectra.[46,68] For 2,2,2–trifluoroethanol, the $T_1$ state of anthraquinone (AQ) has $^3\pi–\pi^*$ character, and the ZFS $|D|$ is 3.1 GHz,[68] while in n–octane the $T_1$ state of AQ is a $^3n–\pi^*$ state, and the ZFS $|D|$ parameter is 9.2 GHz.[69] Based on the $|D|$ value, which is close to 3.1 GHZ for AQ-1,8-NHBu and AQ-1,5-NHBu (Table 2), and is similar to that of the xanthone (the $T_1$ state is $^3\pi–\pi^*$, $|D|$ = 3.3 GHZ),[70] we assign the $T_1$ state of **AQ-1,8-NHBu** and **AQ-1,5-NHBu** as a $^3\pi–\pi^*$ state, which is consistent with the results obtained from natural transition orbital (NTO) analysis.

Generally, $T_1$ states with n–$\pi^*$ character have a relatively short triplet lifetime.[71] However, it has also been reported that $T_1$ states with the $\pi–\pi^*$ character may have a short triplet lifetime, especially for derivatives of aromatic carbonyl compounds.[72,73] The properties of the $T_1$ state of AQ are easily affected by halogen atoms attached at α positions.[74] Previously, it was reported that the shorter triplet lifetime of α–halogenated anthraquinone, having a $T_1$ state with $\pi–\pi^*$ character, is attributed to the interaction of the halogens at α positions with the carbonyl group, which may affect the geometry of the lowest triplet state.[71] In the case of **AQ-1,8-NHBu** and **AQ-1,5-NHBu**, we propose that the substitution of nitrogen atoms at α positions might affect the properties of $T_1$ state.

To gain further insights on the triplet state properties of the studied compounds, we performed a computational investigation (see computational methods section in Supporting Information for details). Figure 4 and Figure S6 show the Natural transition orbitals (NTO) contributing to the lowest excited states of **AQ-1,8-NHBu** and **AQ-1,5-NHBu**. Their $S_1$, $T_1$ and $T_2$ excited states are of $\pi–\pi^*$ character. In contrast, the $T_3$ state has a clear n–$\pi^*$ character. The computed energies of the lowest-lying excited states are reported in Scheme 3. As shown in

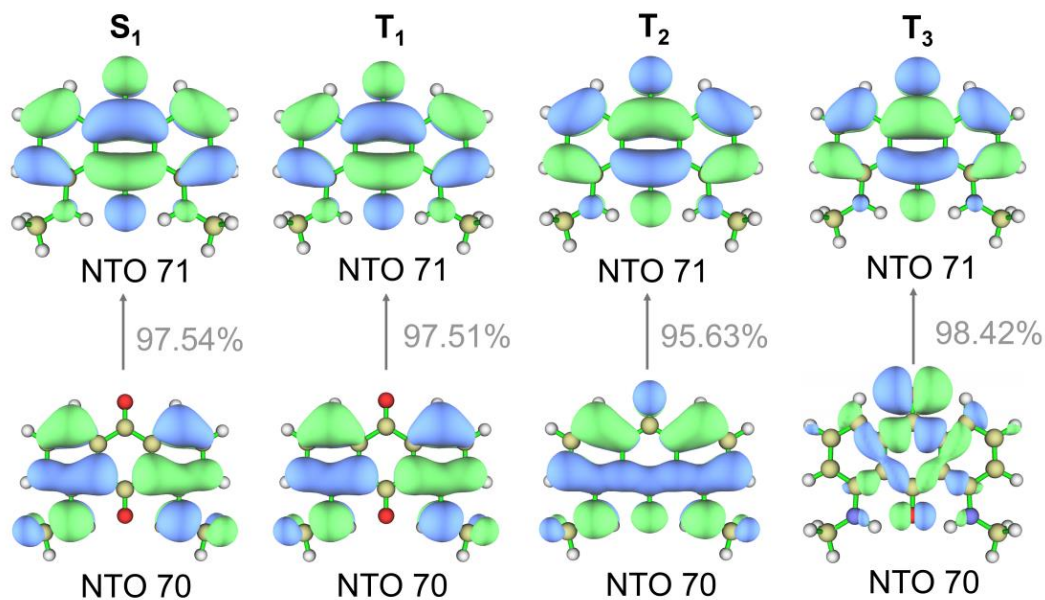

**Figure 4.** Natural transition orbital (NTO) contributing to the $S_1$, $T_1$, $T_2$ and $T_3$ states of **AQ-1,8-NHBu**. Computed at TDA-DFT (PBE0/6-311G (d, p)) level. Isovalue = 0.02 for the map.

Scheme 3, following excitation to $S_1$ only two ISC channels are available at room temperature (RT), i.e., $S_1 \rightarrow T_1$ and $S_1 \rightarrow T_2$. This aligns with the experimental observations of the $S_1 \rightarrow T_2 \rightarrow T_1$ cascade occuring within 0.31 ns in toluene for the two compounds. These ISC channels are notably different from the ones previously reported for anthraquinone derivatives,[75,76] which were postulated to follow El-Sayed rules. To understand why **AQ-1,8-NHBu** and **AQ-1,5-NHBu** still possess very efficient ISC despite the sole π−π* character of their excited states, further computational investigations including the calculation of ISC rate constants at different levels of approximation (see discussion below) were performed.

**Scheme 3. The Proposed Photophysical Process of (a) AQ-1,8-NHBu and (b) AQ-1,5-NHBu** [a]

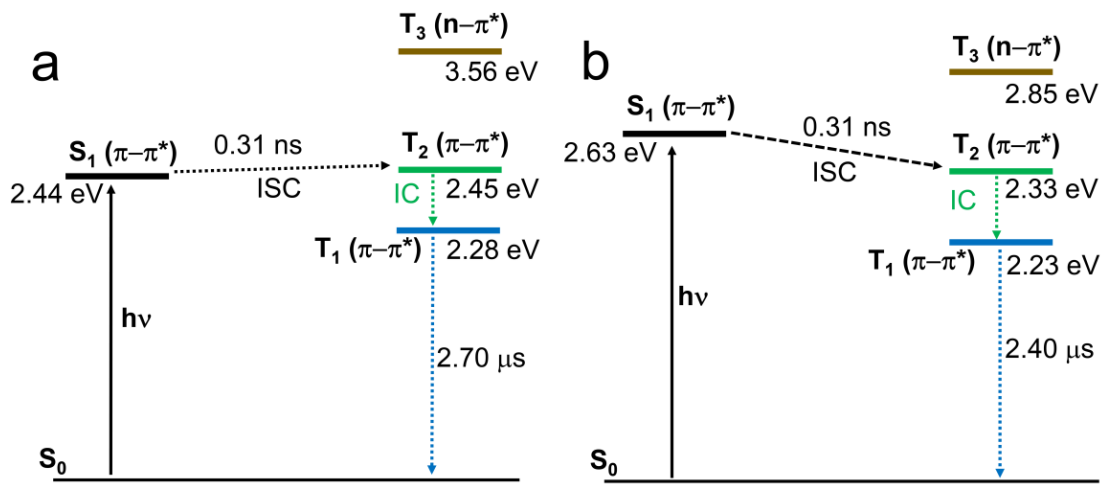

[a] The energy levels of the excited states are calculated at their respective optimized geometries are obtained with SCS-ADC (2)/def2-TZVP.

The ISC rate constant, $k_{if}$, between two electronic states ($\psi_i$, $\psi_f$) can be determined using Fermi's golden rule (FGR) as outlined in eq 1,[77] where $\rho$ represents the density of states, and $\langle \Psi_f | \hat{\mathcal{H}}' | \Psi_i \rangle$ denotes the perturbation matrix.

$$k_{if} = \frac{2\pi}{\hbar} |\langle \Psi_f | \hat{\mathcal{H}}' | \Psi_i \rangle|^2 \rho \qquad (1)$$

Table 3 collects the calculated ISC rate constants at 293.15 K and 80 K for **AQ-1,8-NHBu** and **AQ-1,5-NHBu**. When considering the Franck-Condon (FC) approximation only, ISC rate constants appear significantly smaller (Table S2), not falling within the expected range of $10^8$ to $10^{10}$ s$^{-1}$. Incorporating the Herzberg-Teller like effect, i.e., FC/HT, leads to more reasonable ISC rate constants as compared to the experimental ones. As seen in Table 3, the most favorable ISC channel corresponds to $S_1 \rightarrow T_2$ for **AQ-1,8-NHBu**. Once $T_2$ state is populated rapid internal conversion (IC) to $T_1$ state follows. Finally, $T_1$ state can deactivate by emitting a photon (phosphorescence) or through nonradiative ISC decay back to the ground state. The two compounds behave in a very similar way.

The calculated T$_1$→S$_0$ ISC rate constant for **AQ-1,8-NHBu** amounts to 7.45×10$^6$ s$^{-1}$ at 293.15K, resulting in a T$_1$ lifetime of approximately 0.134 microseconds (Table 3). This aligns closely with the experimental value of 2.7 microseconds obtained from nanosecond transient absorption spectra at 20°C. The agreement between the calculated and measured T$_1$ state lifetime for **AQ-1,5-NHBu** is a bit less good. The discrepancy is attributed to the larger computed displacement (*K*) factor for **AQ-1,5-NHBu** (Table S4), suggesting that the use of the harmonic approximation is less suitable in this specific case.

Finally, to determine the most populated spin sublevel of T$_1$ state, a comparison of the ISC rate constants in the three sublevels was conducted. The zero-field splitting (ZFS) Hamiltonian is given by:[78]

$$\widehat{\mathcal{H}}_{ss} = \hat{S} \cdot D \cdot \hat{S} \tag{2}$$

whereas the full spin Hamiltonian with an extra term about Zeeman effect is given by:[79]

$$\widehat{\mathcal{H}} = \beta H \cdot g \cdot \hat{S} + \hat{S} \cdot D \cdot \hat{S} \tag{3}$$

where $\hat{S}$ is the spin operator, $D$ is the traceless zero-field interaction, $\beta$ is the Bohr magneton, $H$ is the magnetic field, and $g$ is the Landé *g*-factor. It is worth noting that the two Hamiltonians have different eigenfunctions for ZFS and the full spin Hamiltonian. As the magnetic field approaches 0, the relationship between the eigenfunctions of the two Hamiltonians can be stated as follows.

$$|T_x\rangle = \frac{1}{\sqrt{2}}[|-1\rangle - |+1\rangle] \tag{4}$$

$$|T_y\rangle = \frac{i}{\sqrt{2}}[|-1\rangle + |+1\rangle] \tag{5}$$

$$|T_z\rangle = |0\rangle \tag{6}$$

Table 3. Calculated ISC Rate Constants at the FC/HT Approximation of Specific Channels for AQ-1,8-NHBu and AQ-1,5-NHBu at 293.15 K and 80 K.

| Species | ISC Channel | Substate | 293.15 K | | | 80 K | | |
|---|---|---|---|---|---|---|---|---|
| | | | $k_{ISC}$ [a] Components | $k_{ISC}$ [a] Average | $1/k_{ISC}$ [b] | $k_{ISC}$ [a] Components | $k_{ISC}$ [a] Average | $1/k_{ISC}$ [b] |
| **AQ-1,8-NHBu** | $S_1 \rightarrow T_1$ | −1 | $2.00 \times 10^8$ | $1.32 \times 10^8$ | 0.008 | $7.50 \times 10^7$ | $4.93 \times 10^7$ | 0.020 |
| | | 0 | $7.15 \times 10^3$ | | | $5.27 \times 10^3$ | | |
| | | +1 | $1.95 \times 10^8$ | | | $7.29 \times 10^7$ | | |
| | $S_1 \rightarrow T_2$ | −1 | $1.16 \times 10^9$ | $7.62 \times 10^8$ | 0.001 | $6.83 \times 10^8$ | $4.44 \times 10^8$ | 0.002 |
| | | 0 | $1.53 \times 10^5$ | | | $3.86 \times 10^5$ | | |
| | | +1 | $1.13 \times 10^9$ | | | $6.48 \times 10^8$ | | |
| | $S_1 \rightarrow T_3$ | −1 | $8.59 \times 10^5$ | $6.55 \times 10^5$ | 1.526 | $6.58 \times 10^5$ | $5.67 \times 10^5$ | 1.765 |
| | | 0 | $1.88 \times 10^4$ | | | $1.46 \times 10^4$ | | |
| | | +1 | $1.09 \times 10^6$ | | | $1.03 \times 10^6$ | | |
| | $T_1 \rightarrow S_0$ | −1 | $1.12 \times 10^7$ | $7.45 \times 10^6$ | 0.134 | $1.05 \times 10^6$ | $6.97 \times 10^5$ | 1.434 |
| | | 0 | $2.53 \times 10^3$ | | | $5.94 \times 10^2$ | | |
| | | +1 | $1.12 \times 10^7$ | | | $1.05 \times 10^6$ | | |
| **AQ-1,5-NHBu** | $S_1 \rightarrow T_1$ | −1 | $1.09 \times 10^8$ | $7.31 \times 10^6$ | 0.014 | $1.59 \times 10^8$ | $1.07 \times 10^8$ | 0.009 |
| | | 0 | $4.55 \times 10^2$ | | | $1.22 \times 10^3$ | | |
| | | +1 | $1.10 \times 10^8$ | | | $1.63 \times 10^8$ | | |
| | $S_1 \rightarrow T_2$ | −1 | $8.45 \times 10^8$ | $5.62 \times 10^8$ | 0.002 | $1.36 \times 10^9$ | $9.06 \times 10^8$ | 0.001 |
| | | 0 | $5.13 \times 10^3$ | | | $1.46 \times 10^4$ | | |
| | | +1 | $8.42 \times 10^8$ | | | $1.36 \times 10^9$ | | |
| | $S_1 \rightarrow T_3$ | −1 | $2.56 \times 10^6$ | $1.71 \times 10^6$ | 0.585 | $8.38 \times 10^6$ | $3.35 \times 10^7$ | 0.030 |
| | | 0 | $1.06 \times 10^3$ | | | $4.21 \times 10^4$ | | |
| | | +1 | $2.56 \times 10^6$ | | | $9.22 \times 10^7$ | | |
| | $T_1 \rightarrow S_0$ | −1 | $4.09 \times 10^7$ | $2.84 \times 10^7$ | 0.035 | $3.87 \times 10^6$ | $2.86 \times 10^6$ | 0.350 |
| | | 0 | $4.73 \times 10^6$ | | | $8.53 \times 10^5$ | | |
| | | +1 | $3.95 \times 10^7$ | | | $3.85 \times 10^6$ | | |

[a] In $s^{-1}$. [b] In μs.

Therefore, the z-component of the ISC rate constant is equivalent to sublevel 0, while the x and y components of ISC rate constants are a linear combination of sublevels −1 and +1. As seen in Table 3, the z-component of all the computed ISC channels of **AQ-1,8-NHBu** are smaller which leads to a lower population of the $T_z$ level, in agreement with the experimental observations (Table 2). Next, in Table S3, we calcualted the spin-orbit coupling matrix elements (SOCMEs) about **AQ-1,8-NHBu**. The SOCMEs values can provide additional information to understand the experimentally-observed overpopulation of the $T_y$ level with respect to the $T_x$ level. As seen in Table S3, the y-component for the ISC channels has larger SOCMEs than those of the x-component for the ISC channels. This implies that the y-sublevels of the $T_1$ state possess faster ISC rate constant to relax to the $S_0$ state, corresponding to shorter lifetimes. The results are similar for **AQ-1,5-NHBu** (Table S3), although the trend is more evident for **AQ-1,8-NHBu**.

In summary, we found that the triplet state lifetimes of the two bisamino substituted anthraquinone (AQ) derivatives are unusually short (2.06 μs–5.43 μs). Femtosecond transient absorption spectroscopy demonstrated fast ISC (within 200–300 ps) for the two compounds. Pulsed laser excited time-resolved electron paramagnetic resonance (TREPR) spectroscopy shows that the electron spin polarization (ESP) pattern of the triplet states of these compounds is peculiar, as (e,a,e,a,e,a) (e stands for emissive, a stands for enhanced absorption), which is drastically different from the usual ESP pattern (e,e,e,a,a,a) or (a,a,a,e,e,e) of triplet states obtained by the spin–orbit coupling (SOC) in ordinary compounds. Interestingly, the ESP pattern of the TREPR spectrum inverts from the initial (e,a,e,a,e,a) to the (a,e,a,e,a,e) polarization at longer delay time after the laser flash. The special ESP pattern is attributed to the over-populated and rapid decay of the $T_y$ sublevel of the $T_1$ state, and the relative orientation of the zero field splitting (ZFS) principal axes of the $T_1$ state. These conclusions are supported by theoretical investigations, including the calculation of ISC rate constants for the different spin subleveles, which take into account Herzberg–Teller effects. With the simulations of TREPR spectra, we found that the decay rate constant of the $T_y$ sublevel of the $T_1$ state ($k_y = 6.67 \times 10^5$ s$^{-1}$) is 10–fold faster than that of the other two sublevels. ($k_x = 6.67 \times 10^4$ s$^{-1}$ and $k_z = 6.67 \times 10^4$ s$^{-1}$). Our results and analyses are useful for improving the triplet state lifetime of organic compounds, which is beneficial for applications in the field of photodynamic therapy (PDT), photopolymerization, photocatalysis and photo upconversion.

## ■ ASSOCIATED CONTENT

**Ⓢ Supporting Information**

The Supporting Information is available free of charge on the ACS Publications website at DOI: https://pubs.acs.org/doi/10.1021/acs.jpclett.xxxxxx. Femtosecond and nanosecond transient absorption spectra, time-resolved electron paramagnetic resonance (TREPR) spectra, some calculated details about the two compounds.

## ■ AUTHOR INFORMATION

**Corresponding Authors**

**Jianzhang Zhao** − *State Key Laboratory of Fine Chemicals, Frontiers Science Center for Smart Materials, School of Chemical Engineering, Dalian University of Technology, Dalian 116024, P. R. China;*
*orcid.org/0000-0002-5405-6398;* **Email:** *zhaojzh@dlut.edu.cn*

**Daniel Escudero** − *Department of Chemistry, KU Leuven, Celestijnenlaan 200F, 3001 Leuven, Belgium;*
*orcid.org/0000-0002-1777-8578;* **E-mail:** *Daniel.Escudero@kuleuven.be*

**Violeta K. Voronkova** − *Zavoisky Physical-Technical Institute, FRC Kazan Scientific Center of Russian Academy of Sciences, Kazan 420029, Russia;* **Email:** *vor18@yandex.ru*

**Mariangela Di Donato** − *LENS (European Labor NonLinear Spectroscopy), Via N. Carrara 1, 50019 Sesto Fiorentino, Firenze, Italy.*
*orcid.org/0000-0002-6596-7031;* **Email:** *didonato@lens.unifi.it*

**Authors**

***Ruilei Wang*** *– State Key Laboratory of Fine Chemicals, School of Chemical Engineering, Dalian University of Technology, Dalian 116024, P. R. China.*

*orcid.org/ 0000-0001-9104-3126*

***Andrey A. Sukhanov*** *– Zavoisky Physical-Technical Institute FRC Kazan Scientific Center of RAS, Sibirsky Tract 10/7, Kazan 420029, Russia.*

***Yue He*** *– Department of Chemistry, KU Leuven, Celestijnenlaan 200F, 3001 Leuven, Belgium.*

***Aidar Mambetov*** *– Zavoisky Physical-Technical Institute, FRC Kazan Scientific Center of Russian Academy of Sciences, Kazan 420029, Russia.*

**Notes**

The authors declare no competing financial interest.

## ■ ACKNOWLEDGMENTS

J.Z. thanks the National Key Research and Development Program of China (2023YFE0197600), the NSFC (U2001222), the Research and Innovation Team Project of Dalian University of Technology (DUT2022TB10), the Fundamental Research Funds for the Central Universities (DUT22LAB610) and the State Key Laboratory of Fine Chemicals of Dalian University of Technology for financial support. D.E. acknowledges FWO (Project number G079122N) for financial support. A.A.S. and V.K.V. acknowledge financial support from the government assignment for FRC Kazan Scientific Center of RAS. M.D.D. thanks the European Union's Horizon 2020 research and innovation program under grant agreement no. 871124 Laserlab, Europe, for the support.

# ■ REFERENCES

**Supporting Information for:**

# Electron Spin Dynamics of the Intersystem Crossing in Aminoanthraquinone Derivatives: The Spectral Telltale of Short Triplet Excited States

*Ruilei Wang,[a]† Andrey A. Sukhanov,[b]† Yue He,[c]† Aidar Mambetov,[b] Jianzhang Zhao,[a]\* Daniel Escudero,[c]\* Violeta K. Voronkova [b],\* and Mariangela Di Donato[d,e]\**

[a] State Key Laboratory of Fine Chemicals, Frontiers Science Center for Smart Materials, School of Chemical Engineering, Dalian University of Technology, 2 Ling Gong Road, Dalian 116024, P. R. China.
*Email: zhaojzh@dlut.edu.cn

[b] Zavoisky Physical-Technical Institute, FRC Kazan Scientific Center of Russian Academy of Sciences, Kazan 420029, Russia. *E-mail: vio@kfti.knc.ru (V. K V.)

[c] Department of Chemistry, KU Leuven, Celestijnenlaan 200F, 3001 Leuven, Belgium
E-mail: Daniel.Escudero@kuleuven.be

[d] LENS (European Laboratory for Non-Linear Spectroscopy) via N. Carrara 1, 50019 Sesto Fiorentino (FI), Firenze, Italy. *E-mail: didonato@lens.unifi.it

[e] ICCOM-CNR, via Madonna del Piano 10, 50019 Sesto Fiorentino (FI), Italy

† These authors contributed equally to this work.

# Index

# 1. Experimental Section

**1.1. Femtosecond Transient Absorption Spectroscopy.** The femtosecond transient absorption spectra were measured by a setup based on a regenerative amplifier Ti:sapphire laser (Legend, Coherent), pumped by a Ti:sapphire oscillator (Micra, Coherent). The system produces 40 fs pulses at 800 nm, at 1 KHz repetition rate with an average 3.2 W power. The pump beam at 490 nm is obtained by frequency mixing the signal output of a commercial Optical Parametric Amplifier (TOPAS, Light Conversion) with a portion of the 800 nm fundamental laser output, while the white light probe is generated by focusing a small portion of the 800 nm beam on a Calcium fluoride window. The probe beam is split into two parts, acting as probe and reference signal before traversing the sample. Delays are introduced by sending the portion of fundamental radiation used for probe generation through a motorized delay stage. Pump and probe are focused and overlapped at the sample position, the probe then is sent through a monochromator and revealed through a home-made detector. The instrument time resolution is 120 fs. All the data were analyzed with singular value decomposition (SVD) and global fitting, using a linear unidirectional decay scheme.[1]

**1.2. Nanosecond Transient Absorption (ns-TA) Spectroscopy.** The nanosecond transient absorption spectra (ns-TA) were recorded on a LP920 laser flash-photolysis spectrometer (Edin-burgh Instruments, Ltd, UK). All samples were studied under $N_2$ conditions. The sample solution was purged with $N_2$ for 15 min and the cuvette was sealed for measurement. The sample was excited by a 506 nm nanosecond pulsed laser. The data (kinetic decay traces and spectra) were analyzed using L900 software.

**1.3. Time-Resolved Electron Paramagnetic Resonance (TREPR) Spectroscopy.** TREPR measurements were performed with an X-band EPR Elexsys E-580 spectrometer. The sample was dissolved in toluene/2MeTHF (1/3) at 80 K. $O_2$ was removed by a few freeze–pump–thaw cycles. The sample was excited by a pulsed laser with 1 mJ energy per pulse at 506 nm. The EasySpin program package based on MATLAB was used to analyze the data.[2]

**1.4. Computational methods**

Geometry optimizations for the ground and excited states along with frequency calculations were performed with Density Functional Theory (DFT)[3,4] and TD-DFT[5,6] (within the Tamm-Dancoff approximation, TDA[7]) methods; respectively. The PBE0 exchange-correlation functional[8] in combination with the 6-311G(d, p) basis set were used for these calculations. All the above calculations were performed with the Gaussian 16A03[9] package. Next, natural transition orbital (NTO) analyses[10] were performed with Multiwfn 3.7.[11]

Single point second-order algebraic diagrammatic construction ADC(2)[12] calculations were performed on the optimized singlet and triplet geometries, so to obtain accurate singlet-triplet energy splittings and ISC rate constants. The latter calculations were performed with the spin-component-scaled (SCS)[13] approach along with the resolution of identity (RI) approximation and the def2-TZVP basis set[14]. These calculations were performed with Turbomole 7.6.1.[15–18]

ISC rate constants were computed with ORCA 5.0.4[19,20] using the adiabatic Hessian (AH) model. Hessian matrices were obtained from the TDA-DFT calculations, while energies were corrected using the SCS-ADC(2) estimates. We considered temperatures of 80 K and 293 K. In the vibronic model, internal coordinates were used and a Lorentzian broadening (full width at half maximum of 10 cm$^{-1}$) was chosen. These parameters were selected according to previous studies from the literature.[21,22] For the spin-orbit coupling (SOC) calculations, the same exchange correlation functional and basis sets as those of the TDA-DFT calculations was used along with the spin-orbit mean-field approximation (SOMF).[23] For the ISC calculations, both the Franck-Condon and Herzberg-Teller effects were considered, considering three spin substates (MS = (+1, 0, −1)). The ISC calculations made use of a higher DFT integration grid (defgrid3). Finally, the principal axis of the ZFS was also determined from the $D$ tensor using ORCA 5.0.4 and subsequently visualized with Avogadro[24].

## 2. Nanosecond Time-Resolved Transient Absorption Spectra

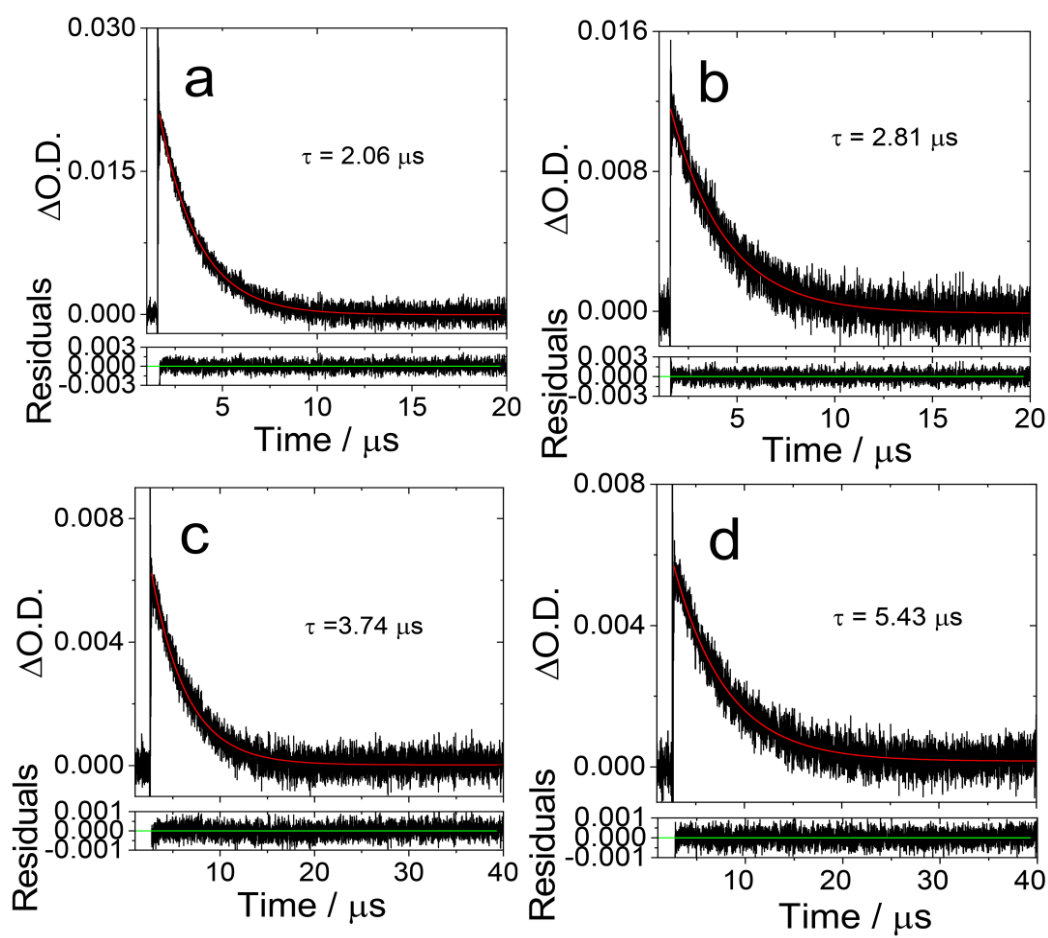

**Figure S1.** The decay traces of **AQ-1,8-NHBu** monitored at 450 nm are presented in deaerated (a) CHX, (b) DCM, (c) ACN, (d) MeOH respectively. $c = 5.0 \times 10^{-5}$ M, $\lambda_{ex} = 506$ nm, 20°C.

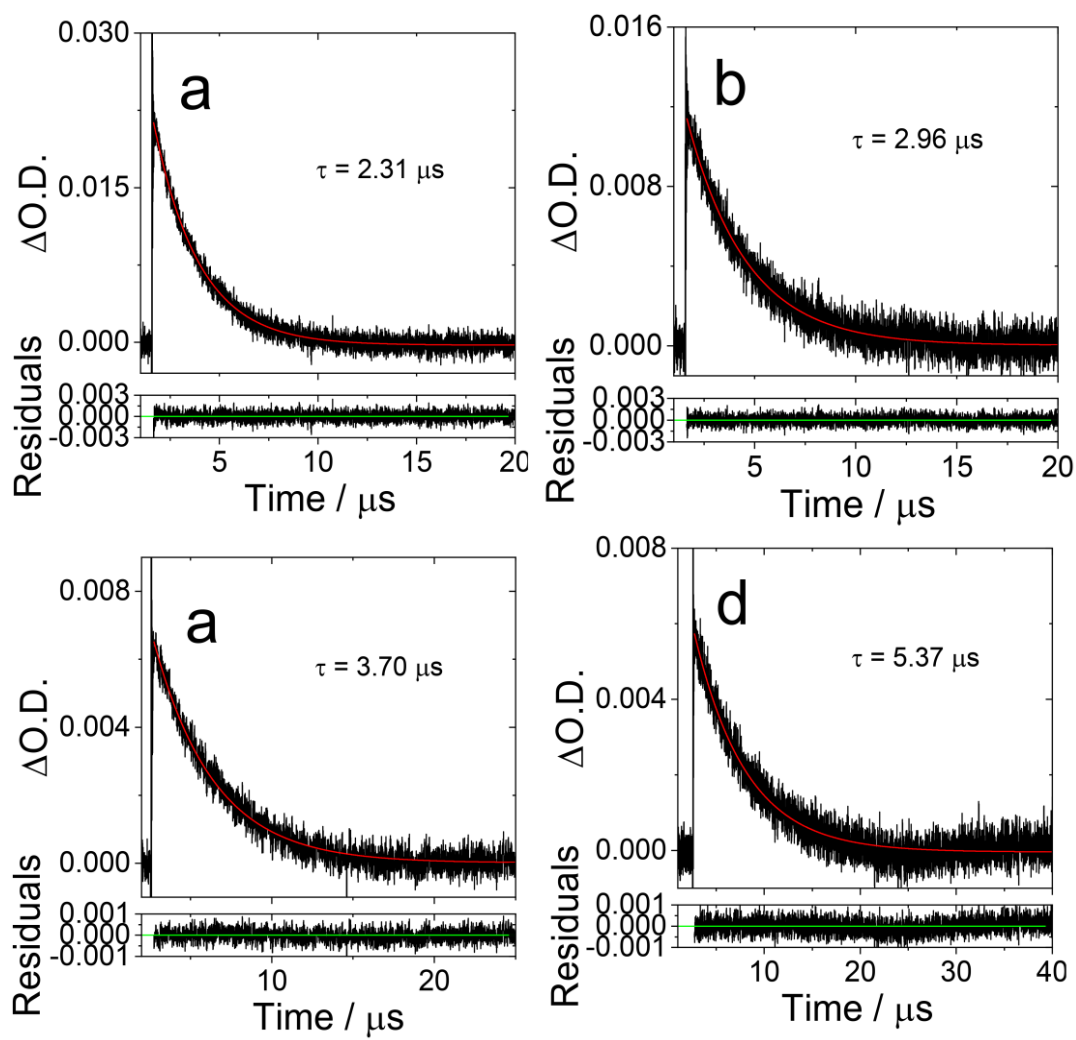

**Figure S2.** The decay traces of **AQ-1,5-NHBu** monitored at 450 nm are presented in deaerated (a) CHX, (b) DCM, (c) ACN, (d) MeOH respectively. $c$ = 5.0 × 10$^{−5}$ M, $\lambda_{ex}$ = 506 nm, 20°C.

## 3. Time-Resolved Electron Paramagnetic Resonance Spectra

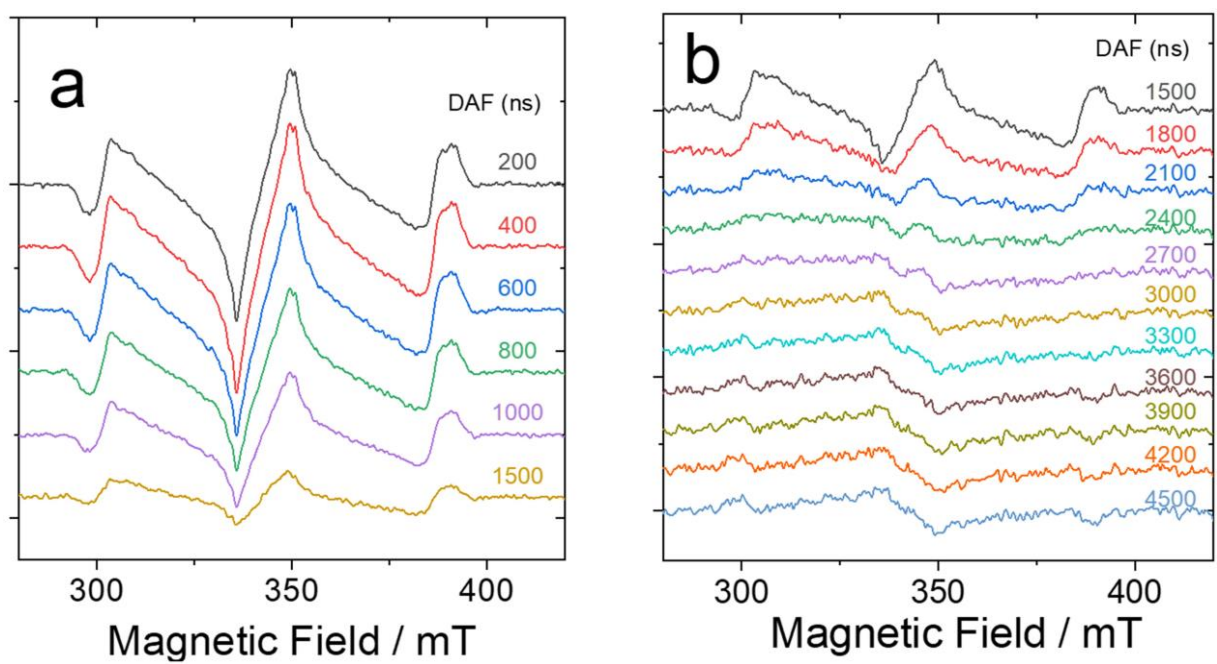

**Figure S3.** TREPR spectra at T = 80 K in X-band of **AQ-1,8-NHBu**. Different delay after laser flash from 200 ns to 4500 ns. Light excitation – 506 nm with energy 1 mJ. $c$ = 1.0×10$^{-4}$ M in toluene/2MeTHF (1/3).

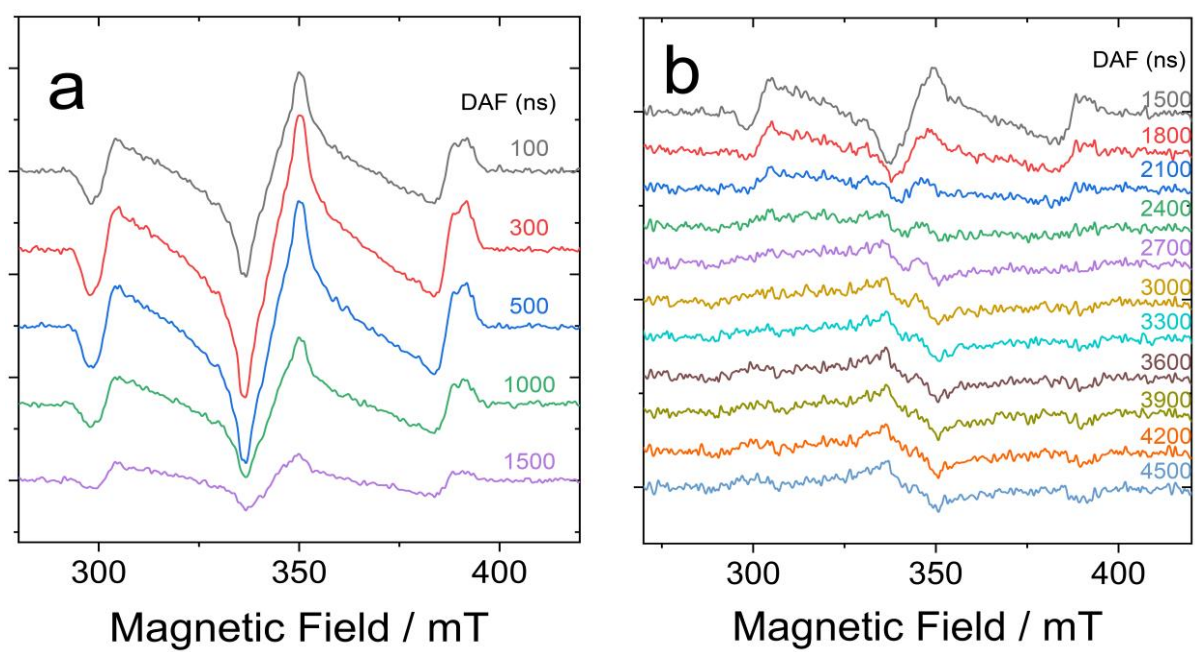

**Figure S4.** TREPR spectra at T = 80 K in X-band of **AQ-1,5-NHBu**. Different delay after laser flash from 200 ns to 4500 ns. Light excitation – 506 nm with energy 1 mJ. $c$ = 1.0×10$^{-4}$ M in toluene/2MeTHF (1/3).

## 4. Femtosecond Time-Resolved Transient Absorption Spectra

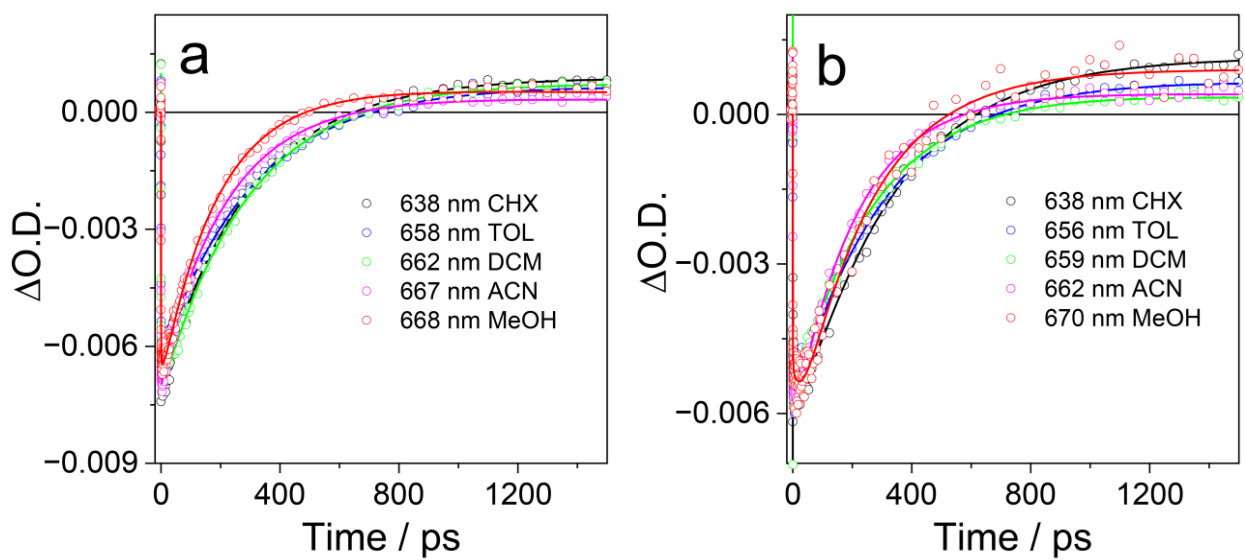

**Figure S5.** Kinetic traces recorded at selected wavelengths for (a) **AQ-1,8-NHBu** and (b) **AQ-1,5-NHBu** in different solvents (CHX, TOL, DCM, ACN and MeOH), scattered points are experimental values, while the continuous lines are the fit obtained from global analysis. $\lambda_{ex}$=490 nm, 20°C.

**Table S1. ISC Rate Constants Retrieved from Global Analysis of the Transient Absorption Spectra Registered for AQ-1,8-NHBu and AQ-1,5-NHBu in Different Solvents.** [a]

| Solvent | AQ-1,8-NHBu [b] | AQ-1,8-NHBu [b] |
|---|---|---|
| Cyclohexane (CHX) | 296.6 | 320.1 |
| Toluene (TOL) | 306.0 | 309.5 |
| Dichloromethane (DCM) | 292.7 | 266.9 |
| Methanol (MeOH) | 189.9 | 210.2 |
| Acetonitrile (ACN) | 226.0 | 241.4 |

[a] $\lambda_{ex}$ = 490 nm. [b] In ps.

## 5. The Calculated Results

**Table S2.** Calculated ISC rate constants at the sole FC approximation of specific channels for AQ-1,8-NHBu and AQ-1,5-NHBu at 293.15 K and 80 K.

| Species | ISC Channel | Substate | 293.15 K $k_{ISC}$ [a] Components | $k_{ISC}$ [a] Average | $1/k_{ISC}$ [b] | 80 K $k_{ISC}$ [a] Components | $k_{ISC}$ [a] Average | $1/k_{ISC}$ [b] |
|---|---|---|---|---|---|---|---|---|
| **AQ-1,8-NHBu** | $S_1 \to T_1$ | -1 | $1.33 \times 10^{0}$ | $1.28 \times 10^{3}$ | $7.82 \times 10^{2}$ | $9.67 \times 10^{-1}$ | $9.26 \times 10^{2}$ | $1.08 \times 10^{3}$ |
| | | 0 | $3.83 \times 10^{3}$ | | | $2.78 \times 10^{3}$ | | |
| | | +1 | $1.33 \times 10^{0}$ | | | $9.62 \times 10^{-1}$ | | |
| | $S_1 \to T_2$ | -1 | $1.13 \times 10^{1}$ | $4.98 \times 10^{4}$ | $2.01 \times 10^{1}$ | $2.85 \times 10^{1}$ | $1.26 \times 10^{5}$ | $7.92 \times 10^{0}$ |
| | | 0 | $1.49 \times 10^{5}$ | | | $3.79 \times 10^{5}$ | | |
| | | +1 | $1.12 \times 10^{1}$ | | | $2.84 \times 10^{1}$ | | |
| | $S_1 \to T_3$ | -1 | $3.86 \times 10^{3}$ | $2.57 \times 10^{3}$ | $3.88 \times 10^{2}$ | $5.35 \times 10^{3}$ | $3.57 \times 10^{3}$ | $2.80 \times 10^{2}$ |
| | | 0 | $5.57 \times 10^{-6}$ | | | $7.72 \times 10^{-6}$ | | |
| | | +1 | $3.86 \times 10^{3}$ | | | $5.35 \times 10^{3}$ | | |
| | $T_1 \to S_0$ | -1 | $1.82 \times 10^{-4}$ | $8.72 \times 10^{2}$ | $1.15 \times 10^{3}$ | $5.17 \times 10^{-5}$ | $2.47 \times 10^{2}$ | $4.04 \times 10^{3}$ |
| | | 0 | $2.61 \times 10^{3}$ | | | $7.42 \times 10^{2}$ | | |
| | | +1 | $1.61 \times 10^{-4}$ | | | $4.57 \times 10^{-5}$ | | |
| **AQ-1,5-NHBu** | $S_1 \to T_1$ | -1 | $2.38 \times 10^{-1}$ | $1.18 \times 10^{1}$ | $8.45 \times 10^{4}$ | $5.47 \times 10^{-1}$ | $2.72 \times 10^{1}$ | $3.68 \times 10^{4}$ |
| | | 0 | $3.50 \times 10^{1}$ | | | $8.05 \times 10^{1}$ | | |
| | | +1 | $2.39 \times 10^{-1}$ | | | $5.49 \times 10^{-1}$ | | |
| | $S_1 \to T_2$ | -1 | $1.42 \times 10^{-1}$ | $9.40 \times 10^{-2}$ | $1.06 \times 10^{7}$ | $3.81 \times 10^{-1}$ | $2.52 \times 10^{-1}$ | $3.97 \times 10^{6}$ |
| | | 0 | $1.15 \times 10^{-5}$ | | | $3.09 \times 10^{-5}$ | | |
| | | +1 | $1.40 \times 10^{-1}$ | | | $3.75 \times 10^{-1}$ | | |
| | $S_1 \to T_3$ | -1 | $7.07 \times 10^{3}$ | $4.62 \times 10^{3}$ | $2.16 \times 10^{2}$ | $2.72 \times 10^{4}$ | $1.78 \times 10^{4}$ | $5.63 \times 10^{1}$ |
| | | 0 | $2.53 \times 10^{1}$ | | | $9.73 \times 10^{1}$ | | |
| | | +1 | $6.77 \times 10^{3}$ | | | $2.60 \times 10^{4}$ | | |
| | $T_1 \to S_0$ | -1 | $8.52 \times 10^{-3}$ | $5.69 \times 10^{-3}$ | $1.76 \times 10^{8}$ | $1.45 \times 10^{-3}$ | $9.70 \times 10^{-4}$ | $1.03 \times 10^{9}$ |
| | | 0 | $8.72 \times 10^{-6}$ | | | $1.49 \times 10^{-6}$ | | |
| | | +1 | $8.55 \times 10^{-3}$ | | | $1.46 \times 10^{-3}$ | | |

[a] In s$^{-1}$. [b] In μs.

**Table S3. SOCMEs of specific ISC channel for AQ-1,8-NHBu and AQ-1,5-NHBu at 293.15 K and 80 K.** [a]

| ISC Channel | Sublevel | AQ-1,5-NH | | AQ-1,8-NH | |
|---|---|---|---|---|---|
| | | 293.15 K | 80 K | 293.15 K | 80 K |
| $S_1 \rightarrow T_1$ | Z | $7.43 \times 10^{-7}$ | $7.43 \times 10^{-7}$ | $9.51 \times 10^{-10}$ | $1.99 \times 10^{-5}$ |
| | X | $2.15 \times 10^{-9}$ | $2.15 \times 10^{-9}$ | $1.80 \times 10^{-1}$ | $1.94 \times 10^{-14}$ |
| | Y | $7.35 \times 10^{-9}$ | $7.35 \times 10^{-9}$ | $3.88 \times 10^{0}$ | $1.38 \times 10^{-8}$ |
| $S_1 \rightarrow T_2$ | Z | $1.13 \times 10^{-9}$ | $1.13 \times 10^{-9}$ | $2.28 \times 10^{-4}$ | $2.28 \times 10^{-4}$ |
| | X | $2.66 \times 10^{-9}$ | $2.66 \times 10^{-9}$ | $2.84 \times 10^{-13}$ | $2.84 \times 10^{-13}$ |
| | Y | $2.19 \times 10^{-9}$ | $2.19 \times 10^{-9}$ | $3.62 \times 10^{-8}$ | $3.62 \times 10^{-8}$ |
| $S_1 \rightarrow T_3$ | Z | $1.57 \times 10^{-3}$ | $1.54 \times 10^{-3}$ | $1.94 \times 10^{-10}$ | $1.94 \times 10^{-10}$ |
| | X | $1.87 \times 10^{-1}$ | $1.75 \times 10^{-1}$ | $2.78 \times 10^{-1}$ | $2.78 \times 10^{-1}$ |
| | Y | $6.69 \times 10^{-1}$ | $6.32 \times 10^{-1}$ | $1.59 \times 10^{-9}$ | $1.59 \times 10^{-9}$ |
| $T_1 \rightarrow S_0$ | Z | $2.00 \times 10^{-6}$ | $2.00 \times 10^{-6}$ | $3.11 \times 10^{-3}$ | $3.11 \times 10^{-3}$ |
| | X | $1.30 \times 10^{-9}$ | $1.30 \times 10^{-9}$ | $3.53 \times 10^{-11}$ | $3.53 \times 10^{-11}$ |
| | Y | $5.14 \times 10^{-9}$ | $5.14 \times 10^{-9}$ | $3.85 \times 10^{-10}$ | $3.85 \times 10^{-10}$ |

[a] In cm$^{-1}$

**Table S4. K*K value of specific ISC channel for AQ-1,8-NHBu and AQ-1,5-NHBu.** [a]

| ISC Channel | AQ-1,8-NHBu | AQ-1,5-NHBu |
|---|---|---|
| | $K * K$ | $K * K$ |
| $S_1 \rightarrow T_1$ | 0.68699 | 2.57366 |
| $S_1 \rightarrow T_2$ | 0.51734 | 1.44534 |
| $S_1 \rightarrow T_3$ | 2.41006 | 4.13221 |
| $T_1 \rightarrow S_0$ | 1.75466 | 2.2807 |

[a] Possibility of errors in the vibronic calculations

**Table S5. The Computed Zero Field Splitting Parameters ($D$ and $E$).** [a]

| | $D$[b] | $E$[b] |
|---|---|---|
| **AQ-1,8-NHBu** | −1114 | −285 |
| **AQ-1,5-NHBu** | −1219 | −262 |

[a] Only under the condition of spin-spin interaction. [b] In MHz.

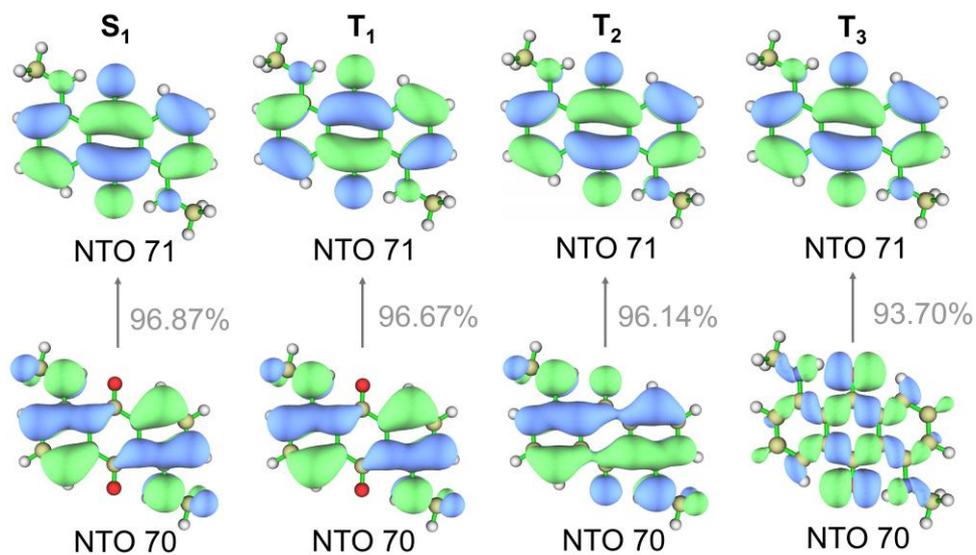

**Figure S6.** Natural transition orbital (NTO) contributing to the $S_1$, $T_1$, $T_2$ and $T_3$ states of **AQ-1,5-NHBu** (TDA-DFT (PBE0/6-311G (d, p))) (isovalue = 0.02)